# Examining socioeconomic factors to understand the hospital case-fatality rates of COVID-19 in the city of São Paulo, Brazil

Camila Lorenz[1*], Patricia Marques Moralejo Bermudi[1], Marcelo Antunes Failla[2], Breno Souza de Aguiar[2], Tatiana Natasha Toporcov[1], Francisco Chiaravalloti Neto[1#] and Ligia Vizeu Barrozo[3#]

1 Departamento de Epidemiologia, Faculdade de Saúde Pública da Universidade de São Paulo, São Paulo, SP, Brazil

2 Gerência de Geoprocessamento e Informações Socioambientais (GISA) da Coordenação de Epidemiologia e Informação (CEInfo) da Secretaria Municipal de Saúde de São Paulo, SP, Brazil

3 Departamento de Geografia, Faculdade de Filosofia, Letras e Ciências Humanas e Instituto de Estudos Avançados da Universidade de São Paulo, São Paulo, SP, Brazil

\# These senior authors contributed equally.

\* Corresponding Author: C. Lorenz, Department of Epidemiology, School of Public Health, University of Sao Paulo, Brazil. E-mail: cammillalorenz@gmail.com

**Financial Support:** This work was supported by the Conselho Nacional de Desenvolvimento Científico e Tecnológico (CNPq) [grant numbers 301550/2017-4 to LVB and 306025/2019-1 to FCN]; and the São Paulo Research Foundation (FAPESP) [grant numbers 2017/10297-1 to CL and 2020/12371-7 to PMMB].

**ABSTRACT**

Understanding differences in hospital case-fatality rates (HCFRs) of coronavirus disease (COVID-19) may help evaluate its severity and the capacity of the healthcare system to reduce mortality. We examined the variability in HCFRs of COVID-19 in relation to spatial inequalities in socioeconomic factors across the city of São Paulo, Brazil. We found that HCFRs were higher for men and for individuals aged 60 years and older. Our models identified per capita income as a significant factor that is negatively associated with the HCFRs of COVID-19, even after adjusting by age, sex and presence of risk factors.

**BACKGROUND**

Coronavirus disease (COVID-19) was declared to be a pandemic by the World Health Organization on March 11, 2020. To date, 9 months later, at least twelve vaccines have been authorized for use, and countries are in various stages of developing national immunization plans. However, these plans will follow different implementation processes, and some countries will take longer to reach their immunization goals. Furthermore, once these goals are achieved, it will take time to determine the levels of protection actually afforded by the vaccine under real-life conditions. Therefore, rather than relying completely on vaccines, risk factors related to the lethality of COVID-19 need to be identified and mitigated. In countries such as Brazil, which has only performed testing at a rate of 11.3/100,000 inhabitants [1], the overall case-fatality rate is overestimated. Although it is impossible to know the real impact of COVID-19 fatalities in Brazil, estimating fatality rates based on confirmed hospital cases allows for the identification of specific at-risk populations. In particular, differences in hospital case-fatality rates (HCFRs) may help evaluate both differences in the severity of the disease across the board and also the capacity of the healthcare system to reduce COVID-19 mortality. Unlike morbidity, which depends on the virus' contagiousness, case-fatality rates are dependent on the virulence of the virus [2]. Viral load, biological risk, population vulnerability (e.g., age structure, prevalence of comorbidities), health system efficiency, healthcare accessibility, and disease detection capacity are the most important factors associated with disease lethality [2]. Thus, the present study aimed to identify socioeconomic factors that are associated with the spatial variability of COVID-19 HCFRs in the city of São Paulo (SP), Brazil.

**METHODS**

This ecological study was based on confirmed COVID-19 hospitalizations, independent of outcome, occurring in SP hospitals from February 27 to November 19, 2020 (Figure 1A). The study population was restricted to hospitalized SP residents registered in the National Influenza Surveillance Information System (Sivep Gripe - extraction date: 24 Nov 2020).

We used spatial analysis methods to compare HCFRs in sociodemographically-distinct regions of SP. The CEInfo, a division of SP's Health Secretariat, geocoded the residential addresses of COVID-19 patients, assigned them to a Human Development Unit (HDU), and created an anonymized database. This database was provided by formal request to the São Paulo Electronic Information System (e-SIC, protocol 53197). As the database was limited to secondary, anonymized data aggregated by HDU, study approval was not required from the Ethics Committee on Research with Human Beings in accordance with Resolution No. 510/2016 of the National Health Council [3].

SP is the capital of the Brazilian state of São Paulo and is the primary city in the largest metropolitan area of the southern hemisphere. We used HDUs, as delineated in the Brazilian Atlas of Human Development [4], as the geographic unit of aggregation. HDUs are homogeneous socioeconomic areas based on information garnered from the 2010 Brazilian Demographic Census [3]. SP contains 1,594 HDUs, 140 of which did not have any reported COVID-19 hospitalizations. These HDUs were merged into the adjacent HDU with which they shared the largest contiguous border, resulting in 1,454 HDUs.

Patient demographic and medical information was obtained from the Sivep Gripe database. Only records with information on patient age, sex, HDU, date of hospitalization, and outcome date were included in the analysis. We calculated the observed sex- and age-specific HCFRs for each HDU, as well as expected deaths, which were derived through indirect age- and sex-standardization [5]. The indirect rates were used as offsets in our geospatial models, allowing us to interpret our results as predicted relative risks (RR), with the COVID-19 HCFR as the baseline.

The observed and expected COVID-19 deaths were linked to HDU shapefiles using the HDU codes. We modeled the number of COVID-19 deaths using spatial models which considered two components of the Besag-York-Mollié (BYM) model, intercept and spatial random effects, to represent the spatial autocorrelation of the dependent variable [6]. The neighborhood relationships between HDUs were constructed using the queen contiguity criterion. We modeled observed counts as a Poisson probability distribution, and ran the models in a Bayesian hierarchical context with integrated nested Laplace approximations (INLA).

We identified HDU-specific socioeconomic variables from the Atlas of Human Development database, limiting the selection to those that were not constructs of original fields. We then used principal component analysis to identify a smaller group to test in the models. For each principal component with an eigenvalue above one, we selected the four covariates with the highest absolute eigenvector values, ran bivariate spatial models considering each covariate against the observed COVID-19 deaths, and retained the covariate from the model with the lowest deviance information criterion (DIC). Before doing this, we transformed covariates with outliers using a square root or logarithm. In addition to the HDU-level socioeconomic fields, we created an individual-level variable for private health insurance and an aggregate indicator for the percentage of patients with one or more of the medical risk factors ("yes" represented at least one risk factor, and "no" represented no risk factor. Risk factors considered: postpartum, chronic cardiovascular disease, chronic hematological disease, Down syndrome, chronic liver disease, asthma, diabetes mellitus, chronic neurological disease, other chronic pneumatopathology, immunodeficiency or immunodepression, chronic kidney disease, and obesity). The final covariates in the models did not present collinearity with each other and were standardized to a mean of 0 and a standard deviation of 1.

We mapped the results of the intercept and random effects models using the posterior mean of the predicted RRs as point estimates, following a similar process for the models adjusted for covariates and for spatial random effects. We considered non-informative priors for the fixed effects and penalized complexity priors for the random effects. All analyses were performed with R software (version 4.0.2, R Foundation for Statistical Computing, Vienna, Austria) with the R-INLA (version 20.03.17) [7] and INLAOutputs packages [8].

**RESULTS**

After excluding records for non-SP residents and hospitals, records for patients with confirmed medical diagnoses other than COVID-19, and 29 records without information for patient age, a total of 44,148 records were available for analysis. For the entire municipality of SP, a total of 12,170 COVID-19 deaths were confirmed, representing an overall HCFR of 27.5%. Men accounted for 55.1% of hospitalizations and 56.4% of deaths and had an HCFR of 28.2% (compared to 26.8% for women). The percentages of hospitalizations and deaths as well as HCFRs increased with age; all values were higher for men except for in the youngest age group. HCFRs varied by age and sex: 4.4% (overall), 4.5% (male) and 4.6% (female) for the 0-19 year age group; 7.2% (overall), 8.3% (male), and 5.8% (female) for the 20-39 year age group; 14.8% (overall), 15.3% (male), and 13.9% (female) for the 40-59 year age group; and 42.1% (overall), 44.3% (male), and 39.7% (female) for the 60 years and older age group.

Figure 1 shows the spatial distribution of the COVID-19 HCFRs, the predicted RRs, per capita income, and rates of the aggregate risk factor variable. Considering the HCFRs (Figure 1A), there was an indication of higher risk in the peripheral regions of the municipality. The RRs, however, demonstrated smoother patterns, and the higher rates in the peripheral regions were more noticeable (Figure 1B). Figure 1C shows higher income in the central regions, and there is a suggestion of an inverse association with HCFR. However, as shown in Figure 1D, the spatial patterns of the comorbidity rates were similar to those of the fatality rates, suggesting a direct association. Table 1 provides the results of the spatial model for COVID-19 HCFRs sorted by HDU. Two covariates were significantly associated with HCFR (95% credible interval) in the adjusted model: per capita income had a negative association (RR = 0.91), and the aggregate risk factor rate had a positive association (RR = 1.11).

**Table 1**. Posterior means of the relative risks and 95% credible interval (CI) for the covariates in the spatial model for hospital case-fatality rate (HCFR) of COVID-19, municipality of São Paulo, state of São Paulo, Brazil, from February 27th to November 19th, 2020.

| Covariate | Relative Risk | 95% CI |
| --- | --- | --- |
| Intercept | 1.03 | 1.01 - 1.05 |
| Per capita income (in log scale) | 0.91 | 0.87 - 0.95 |
| Percentage of registered employees with 18 years or more | 1.00 | 0.97 - 1.03 |
| Percentage of children aged 0 to 5 who do not attend school | 0.98 | 0.96 -1.01 |
| Percentage of people living in households with per capita income less than half the minimum wage and who spend more than an hour to work | 1.00 | 0.97 - 1.02 |
| Percentage of COVID-19 hospitalized people with private care | 0.97 | 0.93 -1.01 |
| Risk factors rate of hospitalized people | 1.11 | 1.06 - 1.15 |

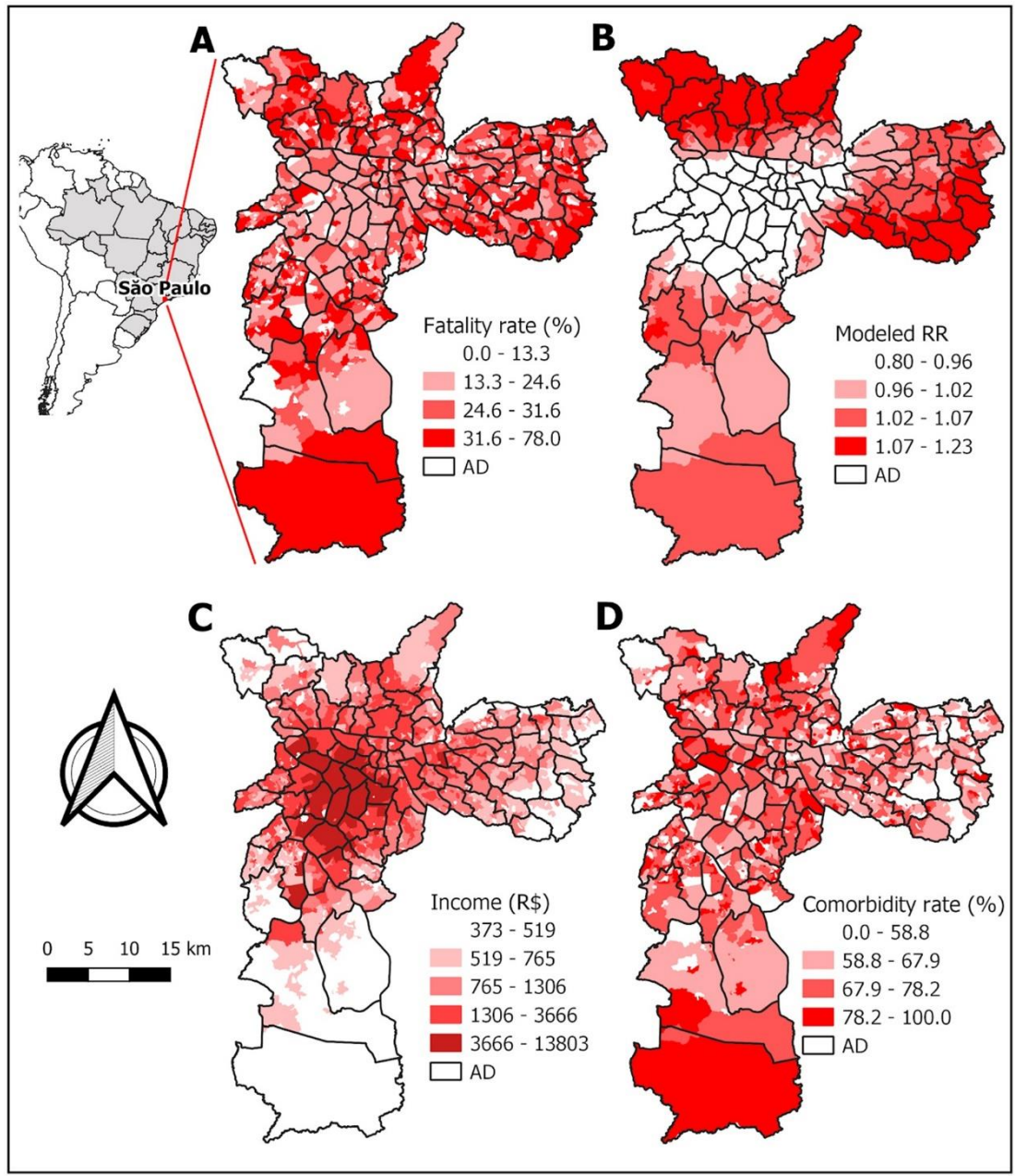

**Figure 1.** Study area and: A) COVID-19 HCFR (%); B) Posterior means of the modelled Relative Risks (RR) for COVID-19 HCFR; C) Per capita Income (in Reais); D) Risk factors rate of hospitalized people. Human Development Units (HDU) of the municipality of São Paulo, state of São Paulo, Brazil, South America, from February 27th to November 19th, 2020. Note: The boundaries of "Administrative District (AD)" are also indicated on the maps as they are widely used by health managers.

## DISCUSSION

The results of the present study indicated an overall HCFR of 27.5% for the metropolitan São Paulo area, similar to results from China for a retrospective cohort of 191 patients at two hospitals [9]. Few studies to date have reported on the main factors associated with fatality rates in hospitalized patients. Regarding age and sex, we found that the fatality rate was higher in men and in those 60 years and older, similar to what was found in other studies [9, 10]. In China, across all age groups, the death rate among confirmed cases was approximately 2.8% for women and 4.7% for men [11]. Even in countries such as Spain (49%) and Switzerland (47%), which have reported fewer infections in men than in women, men accounted for 63% and 62% of deaths, respectively (based on statistics from mid-April, 2020) [11]. It is well known that men and women differ in terms of risk and severity for diseases involving the immune system [12]. Women are disproportionately affected by autoimmune disorders, while men tend to be more susceptible to infectious diseases, both in terms of prevalence and the severity of the disease [13]. The reasons for sex-based differences in COVID-19 are likely multifactorial and include genetics, lifestyle differences, comorbidities, and hormones [10, 12, 13].

Our models found spatial variations in socioeconomic factors, especially per capita income, to be associated with variations in COVID-19 HCFRs across the SP municipal area. This suggests that a reduction in socioeconomic disadvantage could contribute to a decreased risk of COVID-19-related mortality. Corburn et al. [14] suggested measures to protect dwellers of informal urban housing, homeless people, and those living in precarious environments from exposure to COVID-19. Residents of extremely impoverished areas should be prioritized in vaccine distribution policies. A number of studies have offered empirical evidence regarding the association of human health factors, including physical and mental health, with adverse socioeconomic factors, such as poverty, unemployment, and occupational risks, and have shown these to be associated with negative health outcomes [14, 15]. Socioeconomically disadvantaged people are also more likely to have decreased access to healthcare, healthy food, or recreational facilities, a lower level of physical activity, higher use of alcohol and/or tobacco, and less knowledge of healthy-living standards [14].

**CONCLUSION**

The results of the present study indicated that socioeconomic conditions can have a significant relationship with HCFR in large cities such as SP. In particular, our spatial analyses showed that areas of SP with the highest socioeconomic levels have the lowest COVID-19 fatality rates. Enhanced testing, contact tracing, social distancing, and self-isolation are particularly needed in vulnerable communities. Spatial analyses of the implementation of these methods and of disparities in COVID-19 outcomes may help in the development of policies for at-risk populations in geographically-defined areas. Future studies should focus on the identification of these vulnerable groups, in addition to elderly people and front-line healthcare workers, when determining prioritization in vaccination distribution plans.